\documentclass[conference]{IEEEtran}
\IEEEoverridecommandlockouts
\usepackage{cite}
\usepackage{amsmath,amssymb,amsfonts}
\usepackage{graphicx}
\usepackage{textcomp}
\usepackage{xcolor}
\graphicspath{ {pic/} }
\usepackage{float}
\usepackage{array}
\usepackage{tabularx}
\usepackage{array}
\usepackage{balance}
\usepackage{marvosym}
\usepackage{multirow}
\usepackage{array}
\usepackage{enumitem}
\usepackage{booktabs}
\usepackage{algpseudocode}
\usepackage[ruled]{algorithm2e} 

\def\BibTeX{{\rm B\kern-.05em{\sc i\kern-.025em b}\kern-.08em
		T\kern-.1667em\lower.7ex\hbox{E}\kern-.125emX}}
\begin{document}

	\title{Generating Reliable Friends via Adversarial Training to Improve Social Recommendation}

\author{\IEEEauthorblockN{Junliang Yu$^{1}$, Min Gao$^{2}$, Hongzhi Yin$^{1,*}$\thanks{*corresponding author}, Jundong Li$^{3}$, Chongming Gao$^{1}$, and Qinyong Wang$^{1}$}
	\IEEEauthorblockA{\textit{$^{1}$School of Information Technology and Electrical Engineering, The University of Queensland}\\
		\textit{$^{2}$School of Big Data and Software Engineering, Chongqing University}\\
		\textit{$^{3}$Department of Electrical and Computer Engineering, University of Virginia}\\
		\{jl.yu, h.yin1, chongming.gao, qinyong.wang\}@uq.edu.au, gaomin@cqu.edu.cn, and jundong@virginia.edu}	
}
\maketitle

\begin{abstract}
	Most of the recent studies of social recommendation assume that people share similar preferences with their friends and the online social relations are helpful in improving traditional recommender systems. However, this assumption is often untenable as the online social networks are quite sparse and a majority of users only have a small number of friends. Besides, explicit friends may not share similar interests because of the randomness in the process of building social networks. Therefore, discovering a number of reliable friends for each user plays an important role in advancing social recommendation. Unlike other studies which focus on extracting valuable explicit social links, our work pays attention to identifying reliable friends in both the observed and unobserved social networks. Concretely, in this paper, we propose an end-to-end social recommendation framework based on Generative Adversarial Nets (GAN). The framework is composed of two blocks: a generator that is used to produce friends that can possibly enhance the social recommendation model, and a discriminator that is responsible for assessing these generated friends and ranking the items according to both the current user and her friends' preferences. With the competition between the generator and the discriminator, our framework can dynamically and adaptively generate reliable friends who can perfectly predict the current user' preference at a specific time. As a result, the sparsity and unreliability problems of explicit social relations can be mitigated and the social recommendation performance is significantly improved. Experimental studies on real-world datasets demonstrate the superiority of our framework and verify the positive effects of the generated reliable friends.
\end{abstract}


\section{Introduction}
The pervasive use of social media services generates a massive amount of data such as news and social feeds, making current Internet users struggle to find relevant information for their own needs. By modeling the historical data of users such as explicit ratings and implicit feedback, recommender systems can capture users' preferences and free them from this information overload problem. Meanwhile, recommender systems create plenty of revenue for e-business companies like Amazon and Alibaba, since users can easily find the content they are really interested in and would like to pay for. However, as users normally only consume a tiny fraction of items among a sea of options, while numerous new items are continuously being produced, traditional recommender systems often suffer from the thorny problem of data sparsity and hence fail to satisfy the users. 
\par
An effective solution to this dilemma is to transfer knowledge from other fields or platforms and incorporate them into traditional recommender systems. Social relations, particularly, are one type of extremely useful knowledge because people are usually largely influenced by their friends in decision-making \cite{chevalier2006effect}. Based on this, it is natural for the academia and industry to explore the application of social networks to recommender systems, subsequently developing a lot of social recommendation methods \cite{Jamali2009TrustWalker,ma2011recommender,Guo2015TrustSVD,fan2019graph,liu2018social,Ma2009Learning2,wu2019neural}. However, according to a recent survey \cite{Tang2013Social}, social recommendation is not as successful as expected, and sometimes the incorporation of social connections even degrades the recommendation performance. The failure is attributed to the \textit{diversity} and \textit{unreliability} of social connections. For one thing, a vast majority of existing social recommender systems are based on the simple assumption that all the connected users share similar preferences because of the principle of homophily \cite{Mcpherson2001Birds} and hence directly use the explicit social ties for recommendation. But the real situation is rather complex, as the online community is quite different from the offline community in terms of the scale and the possibilities to build connections. Typically, in the online social networks, users can easily expand their friend circles and form different types of relationships such as colleagues, classmates, relatives, among others. Apparently, not all of these social relations have a positive impact on quality-improving for recommendation as most of them may hardly reach a consensus with each other in all the aspects of user preferences. In addition, because of the open nature of social networks, social media users are sometimes mixed with a number of malicious accounts which may pose a threat to accurate preference inference \cite{yu2017hybrid}. Hence, it is highly possible that the direct use of explicit social connections will lead to an unsatisfying result. \par

To tackle this problem, a number of subsequent studies turned attention to extracting salutary relations from explicit social networks \cite{yin2011finding,Wang2016Social,Wang2017Learning,Yu2017A}. However, these attempts ignored the fact that social relations are almost as sparse as the user feedback and the filtered explicit relations are inadequate to make a great contribution to improving recommendation quality. For this reason, a better and feasible alternative is to discover \textit{implicit friends} \cite{Ma2013An,yu2018adaptive} who share similar tastes with the current user but do not have social links with her/him in the social network. In detail, Ma \textit{et al.} \cite{Ma2013An} proposed to use rating profiles to search for implicit friends when social networks are not available and Yu \textit{et al.} \cite{yu2018adaptive} adopted network embedding techniques to uncover implicit friends for each user. But neither of their studies has an adaptive evaluation mechanism to assess the quality of identified implicit friends. The extracted social links are then fed into the recommendation model based on the simple assumption that they are much more dependable in reflecting users' preferences than original explicit friends even if the searching policy may be independent from the recommendation process. Besides, almost all the social recommendation methods simply model the training as a static process and do not consider the change of similarity or proximity between two users during training. In fact, it is inevitable that after several optimization iterations, some relations will not be conducive to the model and even become noises due to the complexity of the models with multiple tuning parameters. Therefore, approaches which can not only uncover reliable friends but also can dynamically assess these relationships should be developed. \par

To this end, in this paper we propose a novel social recommendation framework for Top-\textit{N} recommendation which focuses on reliable friends identification with a dynamic evaluation mechanism. Concretely, as the observed social networks are generally very sparse, we firstly identify a few highly reliable friends as \textit{seeded friends} from both the observed and unobserved social networks, who are quite likely to boost the recommendation performance according to existing research \cite{yu2018adaptive}. And then a framework based on Generative Adversarial Network (GAN) \cite{goodfellow2014generative} is developed to dynamically identify reliable social relations under the supervision of the seeded friends and generate high-quality recommendations. More specifically, the framework is composed of two blocks: a generator component and a discriminator component. The generator is responsible for producing a probability distribution that approximates the real distribution of reliable friends, while the discriminator takes charge of ranking the items with BPR \cite{rendle2009bpr} as the recommendation module and assesses the quality of generated implicit friends for each user, or in other familiar words, discriminate the reliable friends from undependable friends. With the competition between the generator and the discriminator, friends who can better predict the current user's preference are more likely to be incorporated into the training process while those who may mislead the recommendation model are less likely to be generated. It is noteworthy that this identification process is dynamic, which means the generator can adaptively produce friends for the recommendation model according to the real-time feedback of the discriminator, instead of making the probability distribution of the reliable friends unchanged. As a result, it can bring better recommendation performance. \par

However, before enjoying the benefits of this built-in dynamic identification process, we have to overcome the discrete item sampling problem in the generator. As each time a friend and one of its items have to be sampled and fed into the recommendation module, our framework cannot be designed like those GANs applied in image generation since the gradient flow would be blocked, which is one of the major issues when GAN is applied in the discrete data domain. Existing IR models such as \cite{wang2017irgan,wang2018graphgan} bypass the generator differentiation problem by utilizing Policy Gradient strategy \cite{richard99nips}. Unfortunately, the variance of the estimated gradients scales linearly with the number of items, thus increases the volatility of adversarial training, particularly in the case that millions of items are involved \cite{danilo2014stochastic}, which makes Policy Gradient a sub-optimal solution. Unlike Policy Gradient, the recently proposed Gumbel-Softmax \cite{eric2017cate} can approximate categorical samples by adding noises sampled from $\mathit{Gumbel(0,1)}$ to the probability vector produced by the generator, which is known as the reparameterization trick. To enable the framework to be trained via back-propagation, we integrate two Gumbel Softmax layers into our framework to bridge the generator and discriminator and simulate the sampling procedure. Owing to the benefit of end-to-end training, our framework can outperform other recommendation models based on discrete adversarial training.\par

Overall, the major contributions of this paper are summarized as follows:
\begin{itemize}
	\item We formally define a social recommendation framework with a built-in dynamic quality-assessing mechanism for social relationships, which, to the best of our knowledge, has not been investigated before. 
	\item We adopt Gumbel Softmax to simulate the friend and item sampling, making our framework trainable with back-propagation, which has been seldom explored in recommender systems.
	\item We introduce adversarial learning to social recommendation to alleviate the problems of sparsity and unreliability of explicit social relations. 
	\item We conduct experiments on multiple real-world datasets to demonstrate the superiority of the proposed social recommendation framework.
\end{itemize} 
The remainder of this paper is organized as follows. We introduce the related work in Section II. The proposed social recommendation framework is illustrated in Section III. The experimental studies are presented in Section IV. We conclude the whole paper and discuss the future work in Section V.

\section{Related Work}
In this section, we briefly review related work on two aspects: social recommendation and adversarial learning in recommender systems.
\subsection{Social Recommendation}
The early exploration on social recommendation mostly focused on how the explicit social relations can be utilized to improve recommendation performance. A wide range of studies mainly and simply assume that explicitly connected users are supposed to share similar preferences due to the principle of homophily. Among them, SoRec \cite{ma2008sorec} and TrustMF \cite{yang2017social} co-factorize the rating matrix and the relation matrix by sharing a common latent space in which purchase and social information meet with. STE \cite{Ma2009Learning} puts forward an ensemble thought that considers user's essential preference as the linear combination of its own explicit preference and those of its friends. Another fusing strategy, social regularization \cite{ma2011recommender} minimizes the gap between the taste of a user and the average taste of its friends through weighted social regularized terms. Later on, some researchers also developed a few models that explain the social influence on users' feedback from different perspectives. A few works \cite{chen2018modeling,xiao2017learning} noticed that user exposure to items has a great impact on recommendation, and then social connections are used to help capture users' exposure to items instead of preference, which are deemed to be less restrictive. Besides, social information is also leveraged to model the order of items to be recommended. In particular, Zhao \textit{et al.} \cite{zhao2014leveraging} developed a social Bayesian personalized ranking method that assign higher ranks to items that their friends prefer.
\par
However, as the inferior performance of directly leveraging explicit social information has been reported, many subsequent studies then concentrated on extracting valuable information from social relations and identifying reliable social connections. Guided by the weak tie theory, Wang \textit{et al.} \cite{Wang2017Learning} applied the EM algorithm to differentiate strong ties and weak ties from all social ties. Liu \textit{et al.} \cite{liu2018social} introduced a new concept, essential preference space, to describe the multiple preferences of users in social recommender systems. In addition, based on the intuition that social tie inherently has various facets indicating multiple trust relationships between users, Tang \textit{et al.} \cite{tang2012mtrust} proposed to discern multi-faceted trust in search of experts of different types. Bao \textit{et al.} \cite{bao2014leveraging} proposed to decompose the original single-aspect trust information into four general trust aspects and employ the support vector regression to incorporate the multi-facets trust information into the probabilistic matrix factorization model. Enlightened by the success of network embedding techniques, IF-BPR \cite{yu2018adaptive} employs a heterogeneous network embedding method \cite{perozzi2014deepwalk} to discover reliable unobserved user connections. Besides, a few other studies proposed different trust metrics to statically weigh the quality of social ties by computing and predicting trust scores between users based on interactions \cite{fazeli2014implicit,taheri2017extracting}. 
\begin{figure*}[t]
	\centering
	\includegraphics[width=\textwidth]{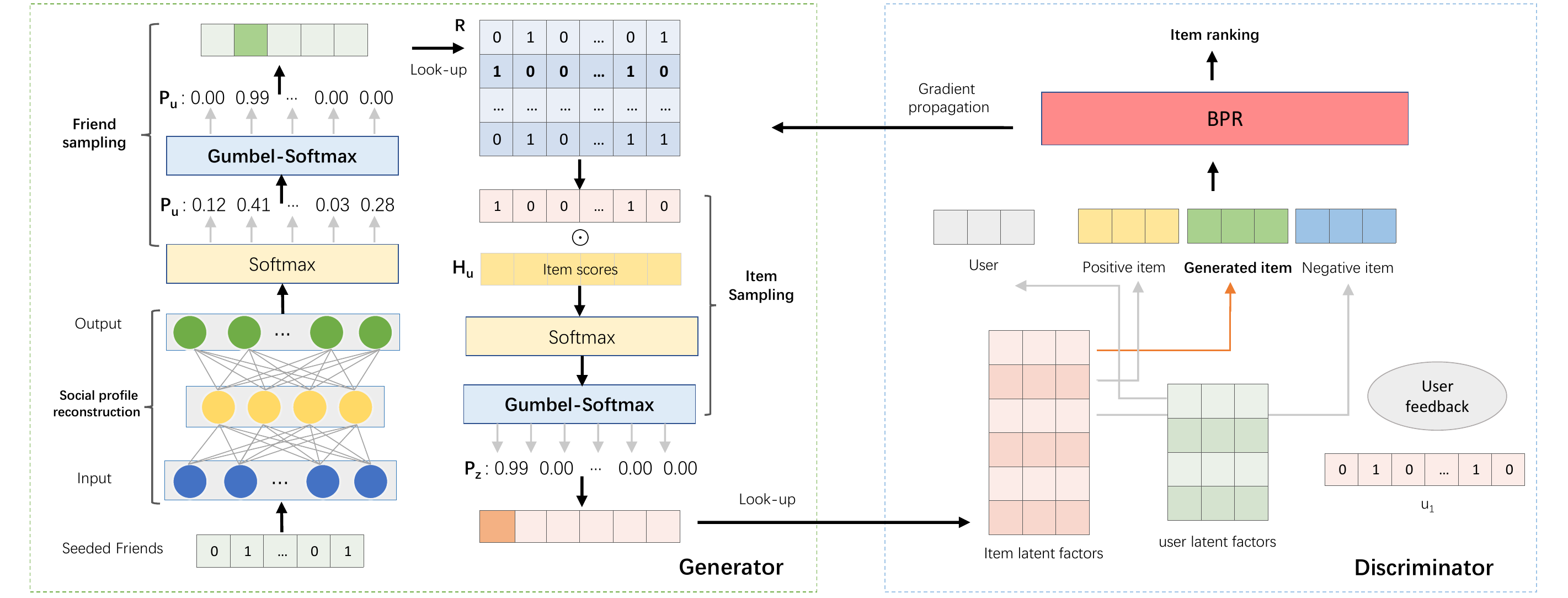}
	\caption{An overview of the proposed RSGAN framework. The generator outputs a probability distribution over a set of candidate latent friends with the seeded friends as the input. Flowing through two Gumbel-Softmax layers, the probability distribution approximates a one-hot vector, which can be used to simulate the item sampling. The discriminator receives the generated item and outputs a ranked item list as recommendations with user feedback and the generated item as input.}
	\label{figure.1}
\end{figure*}

\subsection{Adversarial Training in Recommender Systems}
Recently, adversarial learning \cite{goodfellow2014generative} has achieved great success in various areas such as computer vision and natural language processing. The main idea of adversarial learning is to simulate a minimax game with the generator attempting to imitate the genuine data distribution while the discriminator aiming to differentiate fake examples from the real data. A few pioneering works \cite{cai2018generative,wang2017irgan,wang2018graphgan,wang2018neural,sun2019apl,chae2018cfgan,he2018adversarial,tang2018adversarial} have explored the adversarial learning in recommender systems. IRGAN \cite{wang2017irgan} is the first influential IR model constructed based on GAN. This model unifies the generative model and the discriminative model by using the generator to select the informative negative examples and the discriminator to discriminate negative samples from positive samples. Inspired by IRGAN, GraphGAN \cite{wang2018graphgan} introduced an alternative of softmax called graph softmax to accelerate training, which can greatly improve the computing efficiency. Wang \textit{et al.} \cite{wang2018neural} proposed an adaptive noise sampler to generate adversarial negative samples for neural memory streaming recommender networks, which also boosts the recommendation performance. In addition to the above models focusing on item sampling, Chae \textit{et al.} \cite{chae2018cfgan} made the first attempt to directly learn user profiles with GAN instead of sampling items to advance the recommendation model. He \textit{et al.} \cite{he2018adversarial}  adopted the thoughts of adversarial examples to recommender systems by adding perturbations to the latent factors of recommendation model, and using adversarial training to lower the risk of over-fitting. Then \cite{tang2018adversarial} further extended APR to multimedia recommendation, which also has been shown effective. Besides, \cite{wang2019minimax,wang2019enhancing} investigated a new application of GAN by generating high-level augmented user-item interactions to improve collaborative filtering methods.

\section{The Proposed Framework}
\subsection{Preliminaries}
Firstly, we introduce some important notations used throughout the paper. Let $\mathbf{U}$ denote the user set and $\mathbf{I}$ denotes the item set. $\mathit{m}$ and $\mathit{n}$ are the sizes of the user set and item set, respectively. As our framework is for Top-\textit{N} recommendation, following the convention, we use $\mathbf{R}^{m\times n}$ to represent the interactions between users and items, in which two types of elements are involved: element $r_{u,i}=1$ indicates that user $u$ like item $i$ and element $r_{u,j}=0$ indicates that user $u$ dislike item $j$ or item $j$ is unknown to user $u$. \par

In our framework, the Bayesian personalized ranking model BPR \cite{rendle2009bpr} functions as the recommendation module, which take charges of modeling the order of candidate items. For each user, with the assumption that items appearing in the observed interactions are supposed to be ranked higher than the unobserved ones in the candidate item list, BPR maximizes the margin between the scores of observed items and unobserved ones during training. The assumption of BPR can be formally presented as follows:
\begin{equation}
f: x_{ui}\succeq x_{uj}, r_{ui}=1, r_{uj}=0,
\end{equation}
where $f$ is the model and $x_{ui}$ is the ranking score of item $i$ for user $u$. Each instance of the training data for BPR is a triple $(u,i,j)$ subjected to the constraint $\{(u,i,j)| r_{ui}=1, r_{uj}=0\}$. The loss function of BPR is defined as:
\begin{equation}
\mathcal{L}_{BPR}=-\sum_{(u,i,j)}\log\sigma(x_{ui}(\Phi)-x_{uj}(\Phi))+\lambda_{\Phi}||\Phi||^{2},
\end{equation} 
where $\sigma$ is the sigmoid function, $\Phi$ is the model parameters (user latent factors and item latent factors) to be learned and $\lambda_{\Phi}$ controls the magnitude of $\Phi$ to prevent overfitting. By minimizing Eq. 2, BPR can finally obtain good parameters and generate decent personalized recommendation for each user.\par

\subsection{Motivation}
As has been mentioned in Section 1, almost all the social recommendation models hold a static view towards the effects of social relations. From our perspective, the real-time evaluation for relations is indispensable during training. Let us draw an analogy between the off-line decision making and the training of the social recommendation model. In the real situations, it is common to see that a person consults his friends for advice and then receives different opinions to an item. At first, he may agree with some of his friends but then reaches a consensus with others because of complex psychological activities. Similarly, during the training, social recommendation models are also evolving. A relation initially can be helpful but then may be detrimental to the model due to the complexity of the underlying model with multiple tuning parameters, which resembles the human psychological activities. Therefore, we believe it is completely necessary to build an evaluation mechanism that can assess the real-time effect of relations into the social recommendation model.

\subsection{Model Overview}
In this section, we will have an overall view of the proposed \textbf{R}eliable \textbf{S}ocial recommendation framework , which is based on the \textbf{G}enerative \textbf{A}dversarial \textbf{N}etwork and named \textbf{RSGAN}. \par
It should be noted that RSGAN is designed according to the assumption that users are interested in those items consumed by their reliable friends, which is also the basis of many other social recommendation models. If those items can be identified and fed into the recommendation model to facilitate training, then the recommendation model will be significantly improved. To discover these items, we firstly have to identify the reliable friends for each user, which is the key problem that RSGAN aims to solve.  As shown in Fig. 1, working as an end-to-end framework, RSGAN has two major components: the generator (denoted as $G_{\theta}$) and the discriminator (denoted as $D_\phi$) with parameters $\theta$ and $\phi$ respectively. The discriminator takes charge of ranking the candidate items and producing a recommendation list for the user while the generator is responsible for generating a reliable friend and then sampling an item consumed by this friend which is also likely to be consumed by the current user to enhance the discriminator. Meanwhile, the discriminator punishes the generated friends if items consumed by them are not helpful for advancing the discriminator, and returns gradients to the generator in order to reduce the probabilities of generating such friends. 

\subsection{Friend and Item Generation}
To guide the generator to predict friends with better quality at any time, we first have to obtain a few of \textit{seeded friends} as the supervisor (ground truth), who are highly probable to be helpful in improving the social recommendation performance. The ways to identify seeded friends can be rather diverse. As shown in \cite{yu2018adaptive}, heterogeneous network embedding techniques are good at extracting useful links from both observed and unobserved social networks. Following their work, we use the same way to identify a fraction of reliable friends over the social network and the user-item bipartite graph, who are referred to as \textit{perfect friends} in \cite{yu2018adaptive}, and label them as the seeded friends in our work. However, the number of seeded friends is quite limited, which is not adequate to make a great contribution to improve the quality of recommendation. Hence, the first challenge is that how we can reconstruct the social profile based on the seeded friends in order to have a wide range of friends to be selected. Showing excellent performance in recovering the user rating profile, we believe CDAE \cite{wu2016collaborative}, which is actually a variant of the fully-connected denoised autoencoders, is also able to reconstruct the user social profile and thus we choose it as the generation block of the generator. \par

After the seeded friends for each user are collected, we then encode them into binary vectors as the incomplete user social profiles. These incomplete user profiles are fed into the generator as the initial input. By taking the seeded friends as positive examples and the missing ones as negative examples, CDAE uses cross-entropy loss to optimize the neurons inside and reconstruct the reliable social profiles, and then the layer of softmax outputs the probability distribution over the whole user set $\mathbf{U}$. The distribution can be formally defined as:
\begin{equation}
P_{u}(v|u,S_u) = \frac{exp(c(v|S_{u}))}{\sum exp(c(v'|S_{u}))},
\end{equation}
where $v$ is a latent friend of the user $u$, $c(\cdot)$ is the output of CDAE and $S_{u}$ denotes the seeded friends of $u$. \par

With the possibility distribution of friends known, intuitively, the user with the highest possibility should be sampled as the reliable friend of the user $u$. However, there are two problems for this kind of choice. Firstly, this user also has high possibility to be chosen next time although the distribution is changing when the optimization proceeds, which may lead to the overfitting of the recommendation model as the items consumed by this user will be sampled a lot of times. Secondly, the discrete sampling procedure is non-differential, which means the training is not end-to-end, resulting in an inferior recommendation performance. A few models \cite{wang2017irgan,wang2018graphgan} tackle the non-differentiation problem by utilizing Policy Gradient strategy \cite{richard99nips}. Unfortunately, the Policy Gradient often causes unstable training and slow convergence due to its high variance of the reward, particularly in the case that millions of items are involved, which will further deteriorate the performance of the model \cite{danilo2014stochastic}. An solution to this issue is to relax the discrete items. Unlike Policy Gradient, the recently proposed Gumbel-Softmax \cite{eric2017cate} can approximate categorical samples by applying a differentiable procedure, which can be used to optimize large-scale models like recommendation models \cite{sun2019apl}. To enable the framework to be trained via back-propagation, we integrate two Gumbel-Softmax layers into our framework to simulate the friend and item sampling procedures. \par

For user $u$, let $g_{u} \in \mathbb{R}^{m}$ denote the noise vector whose elements are randomly drawn from $\mathit{Gumbel(0, 1)}$\footnote{$Gumbel(0,1)$ can be sampled with $g=-log(-log(\mu))$, where $\mu\sim Uniform(0,1)$.}. The differentiable friend sampling procedure is expressed as follows:
\begin{equation}
v=\frac{\exp \left(\left(\log p_{u}+g_{\text { noise }}\right) / \tau\right)}{\sum_{i=1}^{n} \exp \left(\left(\log p_{uv'}+g_{i}\right) / \tau\right)},
\end{equation}
where $v$ is the generated vector that is analogous to a one-hot vector. When the hyper-parameter $\tau$, which is conventionally called temperature, approaches 0, $v$ approximates a one-hot vector, representing the sampled friend. By this reparameterization trick, we have decoupled the sampling procedure from the training process and enabled gradients flowing, but this does not change the behavior of our framework. Meanwhile, the noise brings the sampling procedure some randomness, which helps to avoid the overfitting problem and enhance the model if we can find a balance between the exploitation and exploration of randomness and reliability. \par

After the reliable friend being sampled, likewise, we sample the item consumed by the friend with Gumbel-Softmax. The difference between the friend sampling and item sampling is that the possibility distribution of items is not generated by the model. Technically, we define a matrix $\mathbf{H}\in \mathbb{R}^{m\times n}$ whose elements are averagely initialized to represent the scores of items, which is going to be updated during training. The item sampling can be formally defined as:
\begin{equation}
z = GumbelSoftmax(v^{\top}\mathbf{R}\odot\mathbf{H}_{u}),
\end{equation}  
where $z$ is the analogous one-hot vector that represents the generated item and $\odot$ is the element-wise multiplication used to mask the items that have not been consumed by the sampled friend. Finally, the generated item is delivered to the discriminator to help train the recommendation module. Here we explain that why we do not generate items directly instead of generating a friend at first. In fact, if the number of items is huge, directly generating an item will face a problem that there are millions of candidates, which involve much randomness rather than reliability. By contrast, the number of users is relatively smaller, thus generating a friend under the supervision of the seeded friends and then sampling an item from the limited candidates is more feasible. 

\subsection{Adversarial Training}
In this paper, we focus on improving the Top-\textit{N} recommendation and construct the discriminator with the BPR model \cite{rendle2009bpr}. To incorporate the generated item from the friends, we follow the principle of Social BPR \cite{zhao2014leveraging} and impose a social constraint to the discriminator that users tend to assign higher ranks to items that their friends prefer. Hence, the assumption of the recommendation module of the discriminator is formally defined as:
\begin{equation}
\begin{split}
f: x_{ui}&\succeq x_{uz} \succeq x_{uj}, \\&r_{ui}=1, r_{vz}=1, r_{uz}=0, r_{uj}=0.
\end{split}
\end{equation}
With the extra social constraint, the loss function of the discriminator is transformed as follows:
\begin{equation}
\begin{split}
\mathcal{L}_{D_{\phi}}=-\sum_{(u,i,z,j)}(\log\sigma(x_{ui}(\Phi)-x_{uz}(\Phi))\\
+\log\sigma(x_{uz}(\Phi)-x_{uj}(\Phi)))+\lambda_{\Phi}||\Phi||^{2}.
\end{split}
\end{equation}
Each time a quad $(u,i,z,j)\in \mathcal{R}$ is input as an instance, in which item $i$ is the positive item consumed the current user, item $z$ is the generated item, and item $j$ is the negative item that has not been consumed by the current user.
\par

Given the loss function of the discriminator, the objective of the generator can be formulated as maximizing the following expectation:
\begin{equation}
\begin{split}
\mathcal{L}_{G_{\theta}}= -\mathbb{E}[\log\sigma(x_{ui}-x_{uz}),
z\in \mathbf{R}_{v}, f\sim P_{G_{\theta}}(v|u,S_u)].
\end{split}
\end{equation}
The objective of generator can be interpreted that the generator tries to produce such friends whose purchased items can obtain a score that can be comparable with the scores of the items purchased by the current user. As the objectives of the generator and discriminator are contradictory, according to the theory of GAN \cite{goodfellow2014generative}, we have the objective of RSGAN presented as:
\begin{equation}
\begin{split}
\mathcal{L}_{G_{\theta},D_{\phi}}&=\min_{D_{\phi}}\max_{G_{\theta}}-\mathbb{E}[(\log\sigma(x_{ui}-x_{uz})\\
&+\log\sigma(x_{uz}-x_{uj}))],z\in R_{v}, v\sim P_{G_{\theta}}(f|u,S_u).
\end{split}
\end{equation}
By playing a minimax game, the parameters of the discriminator can be updated via gradient descent:
\begin{equation}
\begin{split}
\Phi \leftarrow \Phi-\alpha \cdot \nabla_{\phi} \mathcal{L}((f_{\phi}(i | u)-f_{\phi}(z | u)\\+(f_{\phi}(z | u)-f_{\phi}(j | u)),
\end{split}
\end{equation}
while the parameters of the generator are updated via gradient ascent:
\begin{equation}
\Theta \leftarrow \Theta+\alpha \cdot \nabla_{\theta} \mathcal{L}\left(f_{\phi}(i | u)-f_{\phi}(z | u)\right).
\end{equation}
With the competition between the generator and discriminator, the training process will reach an equilibrium where the generator can produce highly reliable friends and corresponding items while the discriminator can assign good ranks to all the candidate items. 
\par

Let us look back to the evaluation mechanism mentioned in Section 3.2. It is obvious that the adversarial training plays such a role. We can see the adversarial training as a type of persistent evolution of the generator and the discriminator. When the generated friend does not provide an item that can contribute more to the recommendation module, then the delivered gradients will penalize the generator and make it updated towards reducing the possibilities of generating such a friend. At the same time, the possibility of producing a better substitution increases and the discriminator will benefit from this update. It should be noted that even the seeded friends are likely to be penalized if they have a negative impact on minimizing the loss of the discriminator at a certain time. As a result, there are no '\textit{extremely safe relations}' since the process can be viewed as a dynamic and adaptive evaluation mechanism which assesses the quality of relations according to the real-time performance. In other words, the evaluation mechanism is integrated into the model training rather than independent of it. With this mechanism, the social recommendation model becomes robust and flexible.  Meanwhile, we notice that because of the benefits of applying Gumbel-Softmax, RSGAN becomes fault-tolerant. That is, when the model optimization proceeds and the penalized friend becomes useful to the current user, it still has the possibility to be chosen as Gumbel noises introduce more randomness. Besides, since there are two Gumbel-Softmax layers, penalizing an item only has a limited negative impact on the chances of the friend's other items to be chosen.\par

In this paper, we concentrate on the task of Top-\textit{N} recommendation. Actually, RSGAN is also suitable for other tasks including rating prediction. The thoughts of social regularization \cite{ma2011recommender} and co-factorization \cite{ma2008sorec} can be easily implemented with RSGAN by re-designing the discriminator and keeping the generator unchanged. The instance in this section just sets an example for other applications. Finally, there is an unexpected bonus, which is beyond our initial goal, should be mentioned. That is, RSGAN unifies the social recommendation and social link prediction. Because of the interplay of the generator and the discriminator during the training, the CDAE in the generator eventually shows great performance in the task of social link prediction. We will show this bonus in the experiments. The overall process of the training of RSGAN is fully presented in Algorithm 1. 

\begin{algorithm}[t]
	\caption{The training process of RSGAN}
	\LinesNumbered 
	
	\KwIn{Seeded friends $\mathbf{S}$, user feedback $\mathbf{R}$}
	\KwOut{Recommendation lists and user social profiles.}
	
	Initialize the generator $G_{\theta}$ and the Discriminator $D_{\phi}$ \;
	Pretrain the CDAE in $G_{\theta}$ with the seeded friends\;
	\For {each epoch during the training of $G_{\theta}$}{
		\For {each user $u$}{
			Input the seeded friends $S_{u}$ into the generator\;
			Generates $p_{u}$ over the whole user set\;
			Feed $p_{u}$ into the first Gumbel-Softmax layer\;
			Get a one-hot vector representing the friend $v$\;
			Look-up operation for the items consumed by this friend\;
			Do element product with $\mathbf{H}_{u}$ \;
			Feed the obtained vector into the second Gumbel-Softmax layer\;
			Get a one-hot vector representing the item $z$\;
			Deliver $z$ to the discriminator\;
		}
	}
	\For {each epoch during the training of $D_{\phi}$}{
		\For {each user $u$}{
			Receive the generated item $z$ from $G_{\theta}$\;
			Sample a postive item $i$ and a negative item $j$\;
			Train the BPR model with sampled items\;
			Update $D_{\phi}$ based on Eq. 10 and keep $G_{\theta}$ fixed\;
			Update $G_{\theta}$ based on Eq. 11 and keep $D_{\phi}$ fixed.
		}
	}
\end{algorithm}

	\begin{table}[!htb]
		\small
		\renewcommand\arraystretch{1.0}
		\caption{Dataset Statistics}
		\label{Table:1}
		\begin{center}
			\begin{tabular}{|c|cccc|}
				\hline
				Dataset&\#Users & \#Items &  \#Feedbacks &  \#Relations   \\ \hline
				\hline
				LastFM &1,892 &  17,632 & 92,834 & 25,434  \\
				Douban & 2,848 & 39,586 & 894,887 &35,770 \\
				Epinions&18,163  &37,325 & 374,658 &287,260\\
				\hline
			\end{tabular}
		\end{center}
	\end{table}

		\begin{figure}[t]
	\centering
	\includegraphics[width=.5\textwidth]{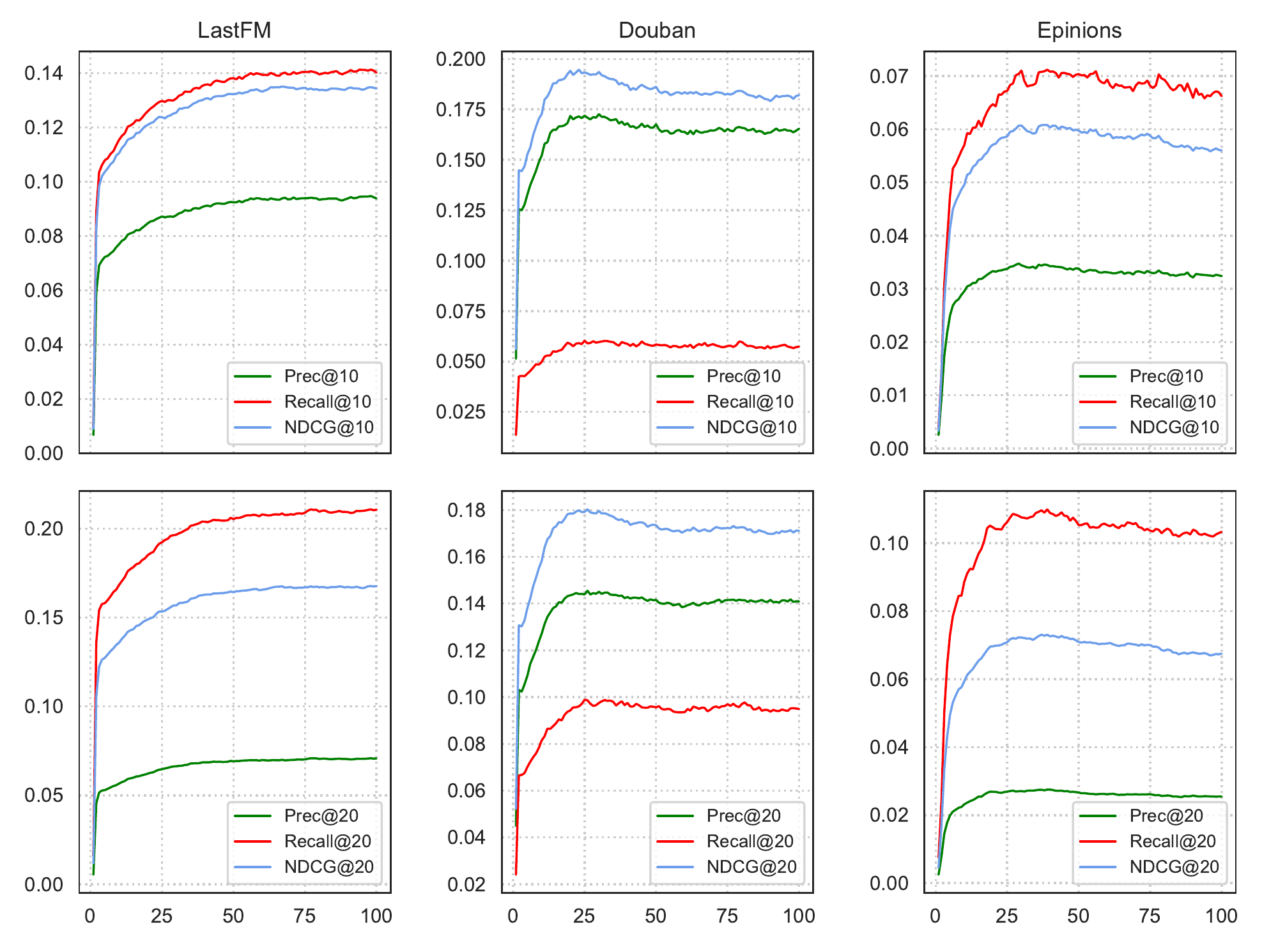}
	\caption{The learning curve of RSGAN}
	\label{figure.2}
\end{figure}

	\begin{table*}[t]
		\centering	
		\caption{Performance comparison of RSGAN and other methods.}
		\label{Table:2}
		\renewcommand\arraystretch{1.10}
		\begin{center}
			\begin{tabular}{|cc|ccccccccc|c|}
				\hline
				Dataset&Metric&Random&BPR&SBPR&TBPR&NeuMF&IF-BPR&IRGAN&CFGAN&RSGAN&Improv.\\ \hline
				\hline
				\multirow{6}{*}{LastFM}
				&Precision@10&5.876\%&7.598\%&8.273\%&8.378\%&8.114\%&9.102\%&8.025\%&8.968\%&\textbf{9.475}\%&4.098\%\\		
				&Recall@10&8.622\%&11.343\%&12.765\%&12.965\%&12.223\%&13.625\%&12.477\%&13.351\%&\textbf{14.128}\%&3.692\%\\			
				&NDCG@10&0.08677&0.10857&0.12195&0.12582&0.12331&0.13043&0.12043&0.12975&\textbf{0.13509}&3.572\%\\
				&Precision@20&4.370\%&5.753\%&6.254\%&6.495\%&6.141\%&6.825\%&6.155\%&6.793\%&\textbf{7.088}\%&3.853\%\\		
				&Recall@20&13.383\%&17.824\%&18.601\%&18.986\%&18.674\%&20.537\%&18.235\%&20.232\%&\textbf{21.091}\%&2.700\%\\			
				&NDCG@20&0.11462&0.13785&0.14664&0.14958&0.14034&0.16252&0.14331&0.16231&\textbf{0.16776}&3.224\%\\
				
				\hline
				\multirow{6}{*}{Douban}
				&Precision@10&11.776\%&13.386\%&15.521\%&16.071\%&16.053\%&16.427\%&15.144\%&15.767\%&\textbf{17.263}\%&5.089\%\\		
				&Recall@10&3.983\%&4.618\%&5.0214\%&5.325\%&5.296\%&5.534\%&4.878\%&5.133\%&\textbf{6.032}\%&8.999\%\\		
				&NDCG@10&0.12361&0.14857&0.17239&0.17883&0.18047&0.18453&0.16463&0.17647&\textbf{0.19469}&5.506\%\\
				&Precision@20&9.374\%&11.530\%&13.005\%&13.269\%&13.648\%&13.884\%&12.276\%&12.633\%&\textbf{14.553}\%&4.818\%\\		
				&Recall@20&6.461\%&7.533\%&8.435\%&8.733\%&8.690\%&8.903\%&8.039\%&8.371\%&\textbf{9.887}\%&11.052\%\\			
				&NDCG@20&0.11208&0.13964&0.16116&0.16447&0.16729&0.16931&0.15623&0.16081&\textbf{0.18030}&6.491\%\\
				
				\hline
				\multirow{6}{*}{Epinions}
				&Precision@10&2.245\%&3.152\%&3.276\%&3.247\%&3.213\%&3.331\%&3.042\%&3.260\%&\textbf{3.477}\%&4.383\%\\			
				&Recall@10&4.983\%&6.135\%&6.355\%&6.418\%&6.224\%&6.785\%&6.033\%&6.627\%&\textbf{7.121}\%&4.952\%\\			
				&NDCG@10&0.04278&0.05423&0.05676&0.05690&0.05538&0.05862&0.05283&0.05787&\textbf{0.06082}&3.753\%\\
				&Precision@20&1.933\%&2.518\%&2.626\%&2.619\%&2.573\%&2.635\%&2.388\%&2.540\%&\textbf{2.750}\%&4.364\%\\		
				&Recall@20&7.956\%&9.729\%&10.343\%&10.413\%&9.918\%&10.438\%&9.577\%&10.310\%&\textbf{10.993}\%&5.317\%\\			
				&NDCG@20&0.04973&0.06553&0.06896&0.06855&0.06702&0.06986&0.06347&0.06937&\textbf{0.07307}&4.595\%\\
				
				\hline
			\end{tabular}
		\end{center}
	\end{table*}

	\begin{figure*}[t]
		\centering
		\includegraphics[width=\textwidth]{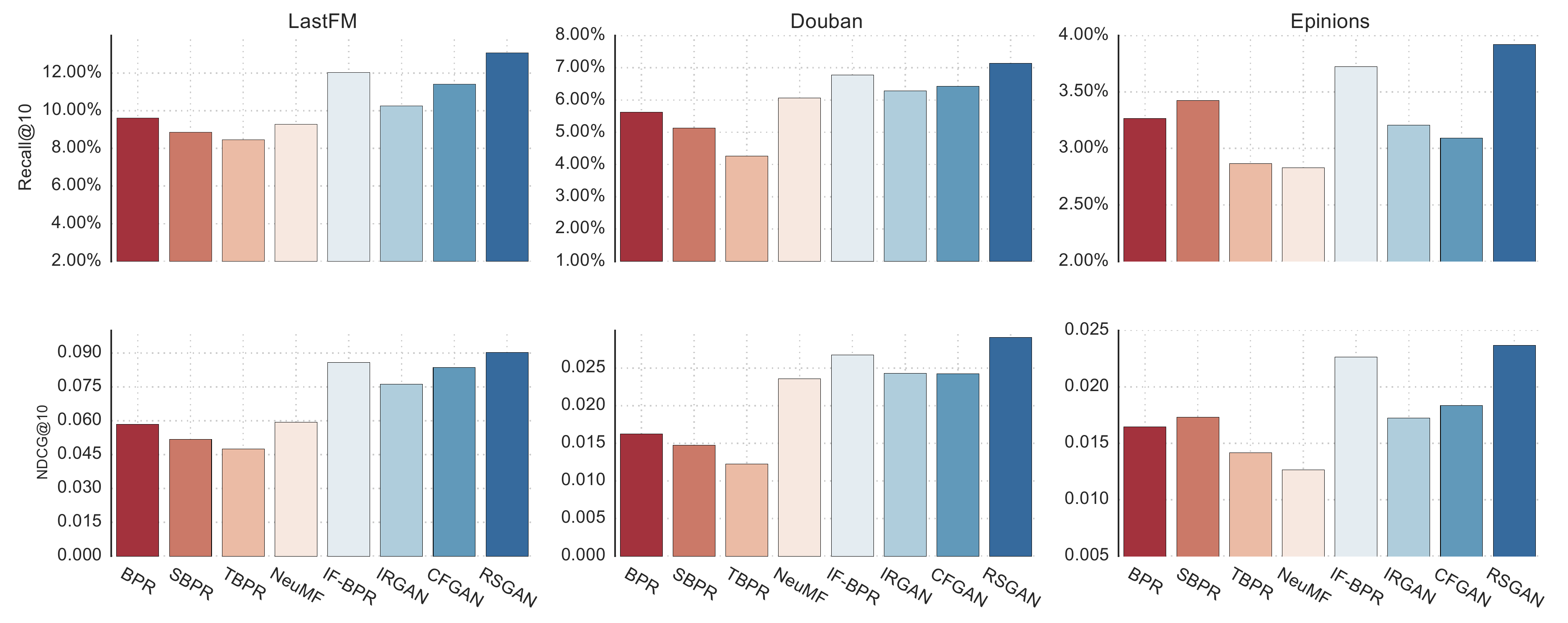}
		\caption{Evaluation on cold-start users with less than 10 feedback.}
		\label{figure.3}
	\end{figure*}
	
	\section{Experiments and Results}
	To evaluate our proposed framework, experiments are conducted to answer the following research questions: (1). Can RSGAN improve the performance of social recommendation?  (2). What are the relevancy and difference between the identified reliable friends and the explicit friends? (3). Can the generator of RSGAN reconstruct the genuine social profiles of the users? 
	
	\subsection{Experimental Settings}
	\noindent\textbf{Datasets.} Three widely used social recommendation datasets, LastFM\footnote{http://files.grouplens.org/datasets/hetrec2011/}, Douban\footnote{http://smiles.xjtu.edu.cn/Download/download\_Douban.html}, and Epinions\footnote{http://www.trustlet.org/downloaded\_epinions.html} are used for experimental evaluations. It should be noted that as the main focus of this paper is to perform Top-$N$ recommendation. So, only the ratings of 4 and 5 in Douban and Epinions are preserved as the user feedback. The statistics of these three datasets are shown in Table 1. For all the datasets, 80\% of the data is kept for training, from which 10\% is selected for validation. Specifically, the parameters of baseline methods are determined by their performance on the validation set. Then the experiments are conducted with 5-fold cross validation and the average performances are presented.
	
	\noindent\textbf{Baseline Methods.} We compare RSGAN\footnote{The implementation of RSGAN: https://github.com/Coder-Yu/RecQ} with these popular item ranking methods: BPR\cite{rendle2009bpr}, SBPR~\cite{zhao2014leveraging}, TBPR~\cite{Wang2016Social}, IF-BPR\cite{yu2018adaptive}, NeuMF \cite{he2017neural}, CFGAN \cite{chae2018cfgan} and IRGAN \cite{wang2017irgan}. Among them, SBPR and TBPR directly use the explicit relations, and IF-BPR identifies implicit friends with network embedding techniques. These three methods are only suitable for social recommendation. BPR and NeuMF are generic models which are shallow and deep respectively. CFGAN and IRGAN are also based on adversarial learning, and comparing RSGAN with them can show the advantage of utilizing social information. Particularly, to verify that the generated friends are the key point, we randomly select 50 users for each user as their random friends and conduct experiments with only the discriminator available and present the results.  
	
	\noindent\textbf{Evaluation Metrics and Configuration.} Two relevance-based metrics - \textit{Precision@K} and \textit{Recall@K}, and one ranking-based metric - \textit{NDCG@K} are used to measure the recommendation performance. For all the models, the regularization coefficient $\lambda_{\Theta}$ is set as 0.001, the batch size is 512, and the dimension of latent factors is 50. For RSGAN, we empirically set the number of sigmoid units in the generator as 200 and the temperature $\tau$ in Gumbel-Softmax as 0.2.

	\subsection{Recommendation Performance}
	We first present the the learning curve of RSGAN in Fig. 2. Table 2 and Fig. 3 present the recommendation performance of all the methods on the whole dataset and on the cold-start user set respectively, and we can make the following observations:
	\begin{enumerate}
		\item In all the cases in Table 2, RSGAN outperforms its counterparts on both the relevance and ranking metrics. Specifically, RSGAN achieves good results in terms of the item ranking and the largest relative improvement (calculated by comparing with the second best performance) is more than 11\%.
		\item The similar results can be observed in the case of recommending for cold-start users as well. RSGAN achieves the best performance while the performance of its main counterparts IRGAN and CFGAN, which is also based on adversarial training, is not very satisfying. We infer that is because IRGAN and CFGAN do not make the most use of the social information and fail to show its capacity in the case of lacking enough user feedback. 
		\item Overall, in these two tests, methods that directly use explicit social links including SBPR and TBPR do not show satisfying performance,  which are in line with the report \cite{Tang2013Social} that the direct use of explicit social relations may have an adverse impact on recommendation performance. Meanwhile, the two implicit friends based methods including IF-BPR and RSGAN show great capacity and beat other methods by a wide margin, which proves that exploiting reliable friends is promising. Besides, the random model with randomly sampled users as friends shows the worst performance in Table 2, which is also a clue to show that it is important to search for relaible friends for social recommendation model. Compared with IF-BPR, RSGAN also gets a decent improvement, we infer it can be attributed to the dynamic evaluation mechanism in RSGAN.
	\end{enumerate}
	
	\begin{figure}[h]
		\centering
		\includegraphics[width=0.5\textwidth]{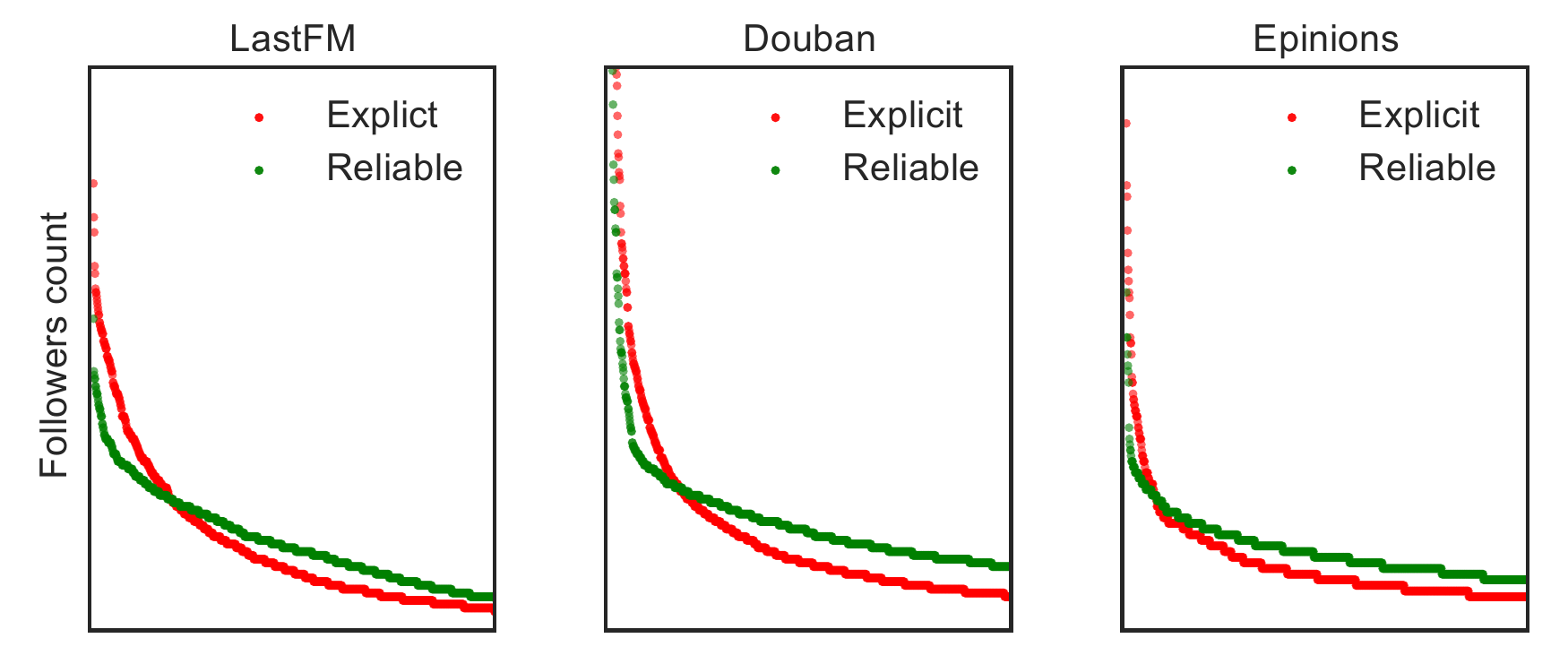}
		\caption{Follower relationship distribution (1000 users are
			randomly selected from each dataset).}
		\label{figure.4}
	\end{figure}
	\subsection{Reliable Friends \textit{vs} Explicit Friends}
	The purpose of this paper is to identify the reliable friends that can perfectly reflect the current user's preference. It is natural to ask what the relevancy and difference between the identified reliable friends and explicit friends are. \par
	To explore the underlying information, firstly, for each user, we sample another 20 users with the highest probabiliies to be chosen as the reliable friends of corresponding users when RSGAN converges. Then we find that about 60\% of seeded friends and about 30\% of explicit friends are preserved in the set of finally identified highly reliable friends. If we consider the final set contains the friends who can best predict the corresponding users' preferences, once again, the findings confirm that the explicit social network is noisy for social recommendation. Besides, we use the reliable links to build a reliable social network and compare the one with the explicit social network in terms of the distribution of followers. Generally, the number of followers of each node in the explicit social network is nearly subject to the power-law distribution. Here in Fig. 4. the distribution of followers of these two types of social networks are drawn. In contrast to the follower distribution of the explicit friends, the one of the reliable friends is much more evenly distributed over the whole crowds. More specifically, if most users follow a small fraction of explicit friends, the social recommendation model is going be less personalized and can not handle the tail users recommendation problem. Besides, users are less likely to meet suprising recommendation originated from the serendipity brought by the diversity of followees, which is like the most-popular recommendation. We believe the follower distribution of reliable links is another reason why RSGAN outperforms other methods.

	\subsection{Social link Prediction}
	Above experimental results have shown the superiority of RSGAN in improving social recommendation. Although RSGAN is not designed to do the task of social link prediction, benefitting from the adversarial training and the user-item interactions involved, the CDAE model integrated into the generator shows great capacity in reconstructing the genuine user social profile, which is beyond our initial expectation and goal. In this section, We compared the strengthened CDAE model with the original CDAE without adversarial learning involved in the training and two widely used link prediction models: DeepWalk \cite{perozzi2014deepwalk} and LINE \cite{tang2015line} in terms of discovering latent social links in this part.\par	
	
	According to Table 3, we can observe that the strengthened CDAE in RSGAN shows great advantages in predicting the possible social links. We believe this is because the preferences to items have a strong impact on the relation building in social recommender systems, and vice versa, according to the theory of homophily. Particularly, with adversarial training, the interplay between the user-item interactions and user-user interactions can be captured, making the CDAE model fullfil its capacity and meanwhile ehance the framework in return.
	 
	\begin{table}[h]
	\centering	
	\caption{Performance of social link prediction}
	\label{Table:3}
	\renewcommand\arraystretch{1.10}
	\begin{center}
		\begin{tabular}{|cc|cccc|}
			\hline
			Dataset&Metric&CDAE&DeepWalk&LINE&RSGAN\\ \hline
			\hline
			\multirow{3}{*}{LastFM}
			&P@10&4.623\%&5.131\%&5.206\%&\textbf{5.587}\%\\		
			&R@10&14.965\%&16.134\%&16.396\%&\textbf{18.271}\%\\		
			&N@10&0.11187&0.12941&0.13181&\textbf{0.14566}\\

			\hline
			\multirow{3}{*}{Douban}
			&P@10&3.931\%&4.265\%&4.133\%&\textbf{4.519}\%\\		
			&R@10&12.825\%&13.558\%&13.407\%&\textbf{14.361}\%\\		
			&N@10&0.09876&0.10635&0.10332&\textbf{0.11479}\\

			\hline
			\multirow{3}{*}{Epinions}
			&P@10&4.664\%&4.125\%&4.385\%&\textbf{5.899}\%\\		
			&R@10&4.206\%&3.988\%&4.073\%&\textbf{5.2675}\%\\		
			&N@10&0.06126&0.05263&0.05457&\textbf{0.07362}\\

			\hline
		\end{tabular}
	\end{center}
\end{table}

	\section{Conclusion}
	Social recommendation suffers from the sparsity and unreliability problem of the explicit social relations. Inspired by the successful applications of GAN in other areas, in this paper, we present a GAN based friend generation framework, named RSGAN, to improve social recommendation.  With the competition between the generator and the discriminator, RSGAN dynamically penalizes the unreliable social relations and adaptively produces reliable friends which can better predict users' preferences. To enable the end-to-end training, we adopt Gumbel-Softmax to relax the discrete sampling, which has been seldom explored in recommender systems. To the best of our knowledge, this is the first work that uses adversarial training to address the unreliability problem of explicit social relations in social recommendation.  Experimental analysis on three real-world datasets demonstrates the superiority of RSGAN and verifies the positive effects of the generated reliable implicit friends. In the future, we will extend RSGAN to other social tasks including rating and link prediction.

\section*{Acknowledgments}
This research is supported by Australian Research Council Discovery Project (Grant No. DP190101985 and DP170103954) and the Fundamental Research Funds for the Central Universities (Grant No. 2019CDXYRJ0011).

	\bibliographystyle{IEEEtran}
	\bibliography{refs}

\end{document}